\documentclass[preprint,aps,nofootinbib]{revtex4}
\usepackage[dvips]{graphicx}
\usepackage{amssymb}
\usepackage{amsmath}
\usepackage{amsmath, amsthm, amssymb, mathrsfs}

\begin{document}

\title{Validity of Emparan-Horowitz-Myers argument in Hawking radiation into  
massless spin-$2$ fields} 

\author{Eylee Jung\footnote{Email:eylee@kyungnam.ac.kr}}
\author{D. K. Park\footnote{Email:dkpark@hep.kyungnam.ac.kr}}

\affiliation{Department of Physics, Kyungnam University,
Masan, 631-701, Korea.}

%\date{\today}

\begin{abstract}
The Hawking radiation for massless spin-$2$ fields is numerically studied 
when the spacetime background 
is $(4+n)$-dimensional Schwarzschild black hole phase. 
In order to check the validity of the Emparan-Horowitz-Myers argument, {\it black holes 
radiate mainly on the brane}, we assume that the radial equation for the massless 
spin-$2$ fields propagating on the brane obeys the master equation approximately. The 
transmission coefficient is computed explicitly by making use of the Hawking-Hartle
theorem.
It is shown that the total emission
rates into the bulk are dominant compared to the rates on the visible
brane when $n \geq 3$. However, the bulk-to-brane relative emissivities per degree of 
freedom always remain $O(1)$ roughly. 
The experimental significance  of these results in the production of 
mini black holes in future colliders is briefly discussed.
\end{abstract}

%\pacs{04.20.Ha, 04.20.Jb, 11.27.+d}

\maketitle

\newpage
One of the most important consequence of the recent brane-world scenarios\cite{bwsc1,bwsc2}
with large or warped extra dimensions is the emergence of low-scale ($\sim 1$ TeV)
quantum gravity. This fact opens a possibility for the copious production of the 
mini black holes in the future colliders such as LHC by high-energy 
collision experiment\cite{hec1,hec2}. In this reason the absorption and Hawking radiation
for the higher-dimensional black holes have been extensively explored recently.

Emparan, Horowitz and Myers(EHM) argued in Ref.\cite{emp00} that the higher-dimensional black
holes radiate mainly on the brane via the Hawking radiation. This argument was supported
numerically in the case of standard model(SM)  field emission by the $(4+n)$-dimensional 
non-rotating black holes\cite{non-rotating1,non-rotating2,non-rotating3}.  
EHM argument was also examined in the higher-dimensional 
rotating black hole background\cite{frol1}.
When black holes have angular momenta, it is well-known that there exist the 
superradiance modes. It was argued that the existence of the superradiance modes may lead
a different conclusion from EHM argument. However, numerical calculation shows
that EHM argument still holds in the scalar emission by $5d$ rotating black holes 
with two different angular momentum parameters\cite{jung05-3}.

In this lettter we would like to re-examine the EHM argument in 
Hawking radiation for massless spin-$2$ fields 
when the spacetime background is $(4+n)$-dimensional Schwarzschild black 
hole whose metric is 
\begin{equation}
\label{metric1}
ds^2 = - h dt^2 + h^{-1} dr^2 + r^2 d \Omega_{n+2}^2
\end{equation}
where $h = 1 - (r_H / r)^{n+1}$ and  the angle part $d \Omega_{n+2}^2$ is a 
spherically symmetric line element in a form
\begin{equation}
\label{angle-part}
d\Omega_{n+2}^2 = d\theta_{n+1}^2 + 
\sin^2 \theta_{n+1} \Bigg[ d\theta_{n}^2 + \sin^2 \theta_n \bigg(
\cdots + \sin^2 \theta_2 \left( d\theta_1^2 + \sin^2 \theta_1 d\varphi^2
           \right) \cdots \bigg) \Bigg].
\end{equation}
Since graviton is not localized on the brane unlike the SM particles, its
emission spectrum may exhibit different behaviors from spectra for other fields.

The emission of the graviton into bulk was numerically  explored 
in Ref.\cite{bulk1,bulk2,bulk3,bulk4,bulk5}. Following the Regge-Wheeler method, it is well-known that 
the gravitational perturbations
in the spacetime dimensions larger than four consist of three modes
according to their tensorial behavior on the spherical section of the background
metric: scalar, vector and 
tensor\cite{reg57-1,reg57-2,reg57-3,reg57-4}. Thus the emission spectra for the 
graviton can be computed via the 
Hawking formula\cite{hawk74-1,hawk74-2}
\begin{equation}
\label{blemi}
\frac{d^2 \Gamma_{BL}}{d \omega dt} =
\frac{(n+1) (n+4)}{2} 
\left[ 2^{n+2} \pi^{(n+3) / 2} \Gamma \left(\frac{n+3}{2}\right) \right]^{-1}
\sum_{A = S,V,T} \frac{\omega^{n+3} \sigma_{A}^{BL}}{e^{\beta_H \omega} - 1}
\end{equation}
where $\beta_H$ is an inverse Hawking temperature, and S, V, and T denote the corresponding
scalar, vector, and tensor modes. 
Of course, $\sigma_A^{BL}$ is an total absorption cross section for each mode.

In 
Ref.\cite{bulk1} the bulk graviton emission rate is compared to those for the SM
fields propagating on the brane and concluded that the bulk graviton emissivities are 
highly enhanced with increasing $n$. However, the bulk-to-brane ratio for the graviton
emissivities was not computed in the paper. Thus, the result of the paper does not lead
any conclusion whether EHM argument is valid or not in the problem of the spin-$2$ field 
emission. In 
Ref.\cite{bulk2} the ratio of bulk graviton emissivity to bulk scalar
was computed, which is summarized in Table I. Table I indicates that the emission of the 
graviton fields become dominant more and more with increasing $n$. 
In this Letter we will compute the brane
decay rates for massless spin-$2$ fields numerically and as a result, 
we will show that the bulk-to-brane ratio of the 
total emissivity becomes $0.76$, $0.66$, $1.59$, $4.25$ and $23.93$ when $n=1$, 
$2$, $3$, $4$ and $6$ respectively. This indicates that the total emissivities into the 
bulk become dominant when $n \geq 3$. Since, however, the bulk spin-$2$ fields have 
$(n+4) (n+1) / 2$ polarization states while brane fields have only two helicities, the 
emission rates per degree of freedom into the bulk become roughly same order with those
on the brane, which strongly support the EHM argument. In spite of $O(1)$ roughly in the
bulk-to-brane relative emissivities the dominance of the bulk emission rates indicates that
we cannot ignore the missing energy portion in the future experiments relating to 
the brane-world black holes.
 
\begin{center}
{\large{Table I}}: Graviton-to-Scalar Ratio in Bulk Emission
\end{center}

\begin{center}
\begin{tabular}{c|cccccc}           \hline 
$n$ & \hspace{.1cm} $0$ \hspace{.1cm} & \hspace{.1cm} $1$ \hspace{.1cm} & 
\hspace{.1cm} $2$ \hspace{.1cm} & \hspace{.1cm} $3$ \hspace{.1cm} & \hspace{.1cm} $4$ 
\hspace{.1cm} & \hspace{.1cm} $6$  \\  \hline \hline
Graviton / Scalar \hspace{.1cm} & \hspace{.1cm} $0.052$ \hspace{.1cm} & \hspace{.1cm} $1.48$ 
\hspace{.1cm} & \hspace{.1cm} $5.95$ \hspace{.1cm} & \hspace{.1cm} $12.1$ \hspace{.1cm} & 
\hspace{.1cm} $18.8$ \hspace{.1cm} 
& \hspace{.1cm} $34.1$ \\ \hline
\end{tabular}
\end{center}

Now we would like to discuss the emission of the spin-$2$ fields on the brane, whose metric
is projected from the $(4+n)$-dimensional Schwarzschild spacetime (\ref{metric1}):
\begin{equation}
\label{metric2}
ds_4^2 = -h(r) dt^2 + h^{-1}(r) dr^2 + r^2 (d\theta^2 + sin^2 \theta \phi^2).
\end{equation}
In fact, the spin-$2$ graviton is a particle living in the bulk. Thus, it seems to
be ridiculous to consider the graviton propagating on the brane. We think this 
bulk-nature of the graviton is a main reason why the radial equations for the 
axial- and polar-perturbation are not uniquely determined\cite{park06-1-1,park06-1-2} 
when the spacetime is 
a projected metric (\ref{metric2}). Since the purpose of this paper is to check the 
validity of the EHM argument in the spin-$2$ field, we will consider the hypothetical
spin-$2$ field whose radial equation is assumed to be obeyed by the master equation as
follows.

In Ref.\cite{non-rotating1,non-rotating2,non-rotating3,ida02} the perturbations for the 
scalar($s=0$), fermion($s=1/2$) and vector($s=1$) fields were discussed in this 
background by employing Newman-Penrose formalism and the following radial master equation was
derived:
\begin{equation}
\label{master}
\Lambda^2 Y + P \Lambda_- Y - Q Y = 0
\end{equation}
where $\Lambda_{\pm} = d / dr_{\ast} \pm i \omega$, $\Lambda^2 = \Lambda_+ \Lambda_-$, 
$d / dr_{\ast} = h \hspace{0.1cm} d / dr$ and 
\begin{equation}
\label{def1}
P = \frac{d}{d r_{\ast}} \ln \left(\frac{r^2}{h}\right)^{-s},
\hspace{1.0cm}
Q = \frac{h}{r^2} \left[ {\cal A}_{\ell s} + (2 s + n + 1) (n s + s + 1) (1 - h) \right]
\end{equation}
with ${\cal A}_{\ell s} = \ell (\ell + 1) - s (s + 1)$.

Although Eq.(\ref{master}) was derived without considering the graviton($s=2$), one can
easily show Eq.(\ref{master}) is valid for the graviton when $n=0$\cite{chandra92}. 
In Ref.\cite{berti03}, furthermore,  Eq.(\ref{master}) is assumed to be valid for arbitrary
positive $n$ for the graviton and derived the physically relevant quasinormal frequencies.
However, one can show by directly applying the Newman-Penrose formalism that the master
equation (\ref{master}) is not valid for the spin-$2$ graviton fields propagating on the 
brane. As commented above, we guess the proper radial equation for the graviton confined
on the brane does not exist due to the bulk-nature of the graviton field. Since, however,
the original purpose of the present paper is to check the validity of the EHM argument of
the spin-$2$ field, we consider the hypothetical spin-$2$ field whose radial equation is 
assumed to be Eq.(\ref{master}) with $s = 2$.   
In this Letter, therefore,  we will use Eq.(\ref{master}) for the computation
of the emission spectra for the spin-$2$ fields propagating on the brane\footnote{Another 
reason for the choice of Eq.(\ref{master}) as a radial equation is as follows. Since
the quasinormal frequencies computed in Ref.\cite{berti03} are physically reasonable,
we can assume that Eq.(\ref{master}) is to some extent valid approximately to describe
the graviton propagating on the brane. In this reason Eq.(\ref{master}) can be used for
the approximate computation of the Hawking radiation into the graviton confined on the 
brane.}. 

Defining $R = f^{-1} Y$ where $f$ is defined as $(1/f) df / dr_{\ast} = -P / 2$, one can 
transform Eq.(\ref{master}) into the Schr\"{o}dinger-like expression $\Lambda^2 R = V_{br} R$
where the effective potential $V_{br}$ is in general complex in the following expression
\begin{equation}
\label{br-poten}
V_{br} = i \omega P + \frac{P^2}{4} + \frac{1}{2} \frac{d P}{d r_{\ast}} + Q.
\end{equation}
Since the method for the computation of the reflection and transmission coefficients is 
explained for the case of the complex potential in Ref.\cite{chandra92}, we will adopt
the procedure of Chandrasekhar 
for the computation of the absorption and emission spectra. The solution 
convergent in the near-horizon and asymptotic regimes are respectively
\begin{eqnarray}
\label{br-solu}
R_{NH}&=& {\cal C}_{NH} r_H^{1 - s} \left(\frac{x_H}{n+1}\right)^{-\rho_n} 
\sum_{N=0}^{\infty} d_{\ell,N} (x - x_H)^{N + \rho_n}
                                                                           \\   \nonumber
R_{\infty}&=& \omega^{-s} x^{s + 1} e^{-i x} \sum_{N = 0}^{\infty} \tau_{N(+)} x^{-(N+1)} + 
{\cal C}_{\infty} \omega^s x^{-s + 1} e^{i x} \sum_{N = 0}^{\infty} \tau_{N(-)} x^{-(N+1)}
\end{eqnarray}
with $x = \omega r$, $x_H = \omega r_H$, $d_{\ell,0} = \tau_{0(\pm)} = 1$ and 
\begin{equation}
\label{factor1}
\rho_n = -\frac{s}{2} - i \frac{x_H}{n+1}.
\end{equation}
The sign for $\rho_n$ is chosen by $\mbox{Im} \rho_n < 0$ to ensure the incoming
behavior of the spin-$2$ wave in the near-horizon regime. The recursion relations
for the coefficients 
$d_{\ell, N}$ and $\tau_{N(\pm)}$ can be explicitly derived by inserting Eq.(\ref{br-solu}) into
the wave equation, {\it i.e.} $\Lambda^2 R = V_{br} R$.

The transmission coefficients $T_{BR}$ for the complex potential (\ref{br-poten}) can be 
derived as follows. Firstly, we relates changes of the
black hole mass($dM$) to changes of the horizon area($d\Sigma$). 
Although this relation is simply
\begin{equation}
\label{m1}
d \Sigma = \frac{4 G_n}{T_H} d M
\end{equation}
for the bulk metric (\ref{metric1}), where $T_H$ and $G_n$ are the Hawking temperature and 
$(4+n)$-dimensional Newton constant, it becomes slightly complicated form for our case as 
follows;
\begin{equation}
\label{m2}
d\Sigma = \frac{\xi_n}{(\pi / \beta_H)^{1 - n)}} d M
\end{equation}
where $\beta_H \equiv 1 / T_H$, 
\begin{equation}
\label{m3}
\xi_n = \frac{2^{5 + 2 n} \pi^2 G_n}{(n+1)^n (n + 2) \Omega_{n+2}}
\end{equation}
and $\Omega_{n+2}$ is an area of a unit $(n+2)$-sphere.

Employing the Hawking-Hartle theorem\cite{hawk72-1,hawk72-2,hawk72-3}
we express the variation in the area in terms of the variations in the spin coefficients.
Finally the variations in the spin coefficients are identified with the perturbation in the 
Weyl tensor, using Ricci identities. Assuming $Y \equiv r h^2 \tilde{R}$ satisfies the master
equation (\ref{master}) where the Weyl tensor $\Psi_0$ is 
$\Psi_0 = \tilde{R}(r) S(\theta) e^{i m \phi - i \omega t}$, one
can derive the transmission coefficient in the following expression
\begin{equation}
\label{tbr1}
T_{BR} = \frac{(n+2) \Omega_{n+2} r_H^{n - 6} \omega^2}{8 \pi G_n \left[\omega^2 + 
         \left(2 \pi / \beta_H \right)^2 \right]}
         | {\cal C}_{NH} |^2.
\end{equation}
Since we adopt the unit $G_0 = 1$, we should know the relation between $G_n$ and 
$G_0$. Let $\gamma_n \equiv G_n / G_0$. Then $\gamma_n$ can be numerically 
computed by noting that $T_{BR}$ in Eq.(\ref{tbr1}) should be saturated
to unity in $\omega \rightarrow \infty$ limit. The numerical result strongly 
suggests that ${\cal C}_{NH}$ is dependent on neither $\omega$ nor $n$, which
implies $\gamma_n = (n+2) \Omega_{n+2} / 8 \pi$.

\begin{figure}[ht!]
\begin{center}
\includegraphics[height=6cm]{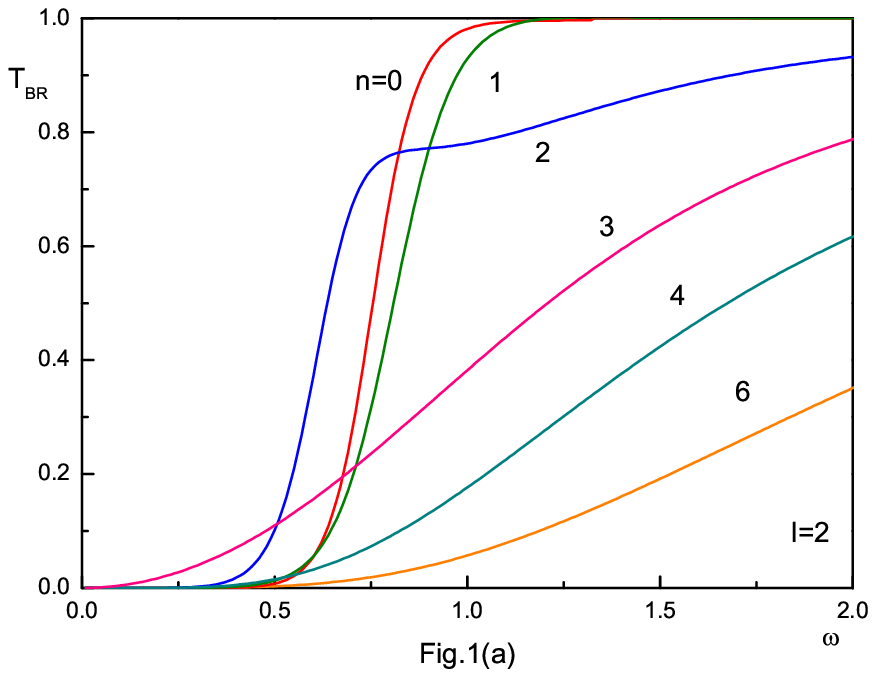}
\includegraphics[height=6cm]{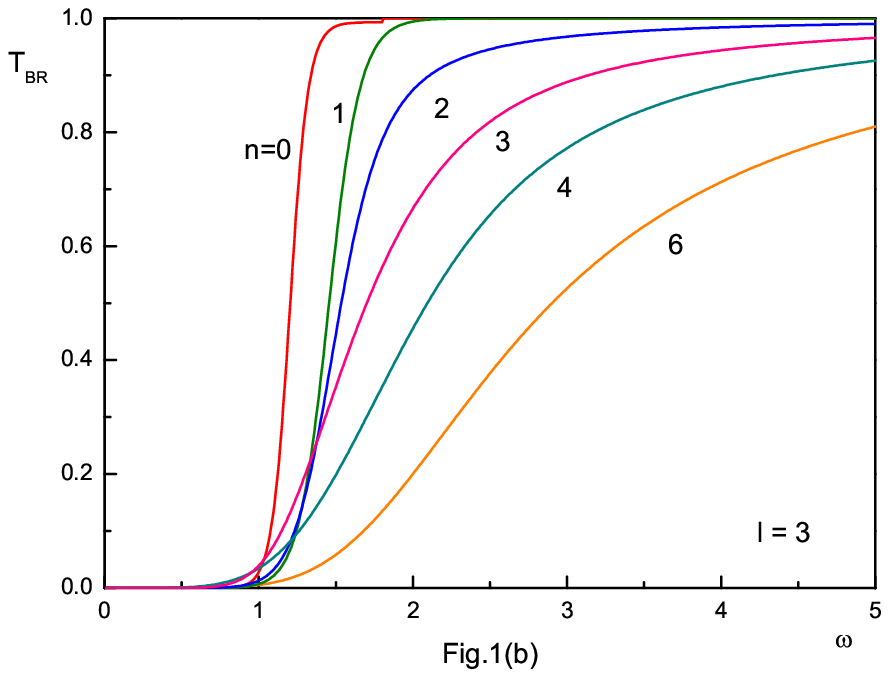}
\caption[fig1]{The $\omega$-dependence of $T_{BR}$ when $\ell = 2$ (a) and 
$\ell = 3$ (b). The increasing rate of $T_{BR}$ tends to reduce with increasing
$n$, which indicates that the barrier heights of the real effective potentials become
higher with increasing $n$.} 
\end{center}
\end{figure}

Fig. 1 is a plot of $T_{BR}$ as a function of the energy $\omega$ for $\ell = 2$
(Fig. 1(a)) and $\ell = 3$ (Fig. 1(b)). As expected $T_{BR}$ is saturated
to unity with increasing $\omega$. The increasing rate of $T_{BR}$ tends to 
decrease with increasing $n$, which implies that the barrier heights
of the real effective potentials become higher with increasing $n$ although we do not know 
the exact expression of the real effective potential. However, the 
low-energy increasing rate of $T_{BR}$ when $\ell = n = 2$ (see Fig. 1(a)) seems to
be extra-ordinarily large, which enables us to guess that the width of the 
potential barrier in this case may be narrower compared to the other cases.

Once $T_{BR}$ is computed, it is straightforward to compute the emission 
spectrum\cite{hawk74-1,hawk74-2}
\begin{equation}
\label{bremi}
\frac{d^2 \Gamma_{BR}} {d \omega d t} = 
\frac{\omega^3 \sigma^{BR}}{\pi^2 \left(e^{\beta_H \omega} - 1 \right)}
\end{equation}
where $\sigma^{BR}$ is a total absorption cross section defined $\sigma_T = \sum_{\ell}
\pi (2 \ell + 1) T_{BR} / \omega^2$. We compute $\sigma^{BR}$ 
numerically by making use of the 
quantum mechanical scattering theories with numerical analytic continuation, 
which was introduced in detail in Ref.\cite{san78,jung05-3}.

\begin{figure}[ht!]
\begin{center}
\includegraphics[height=6cm]{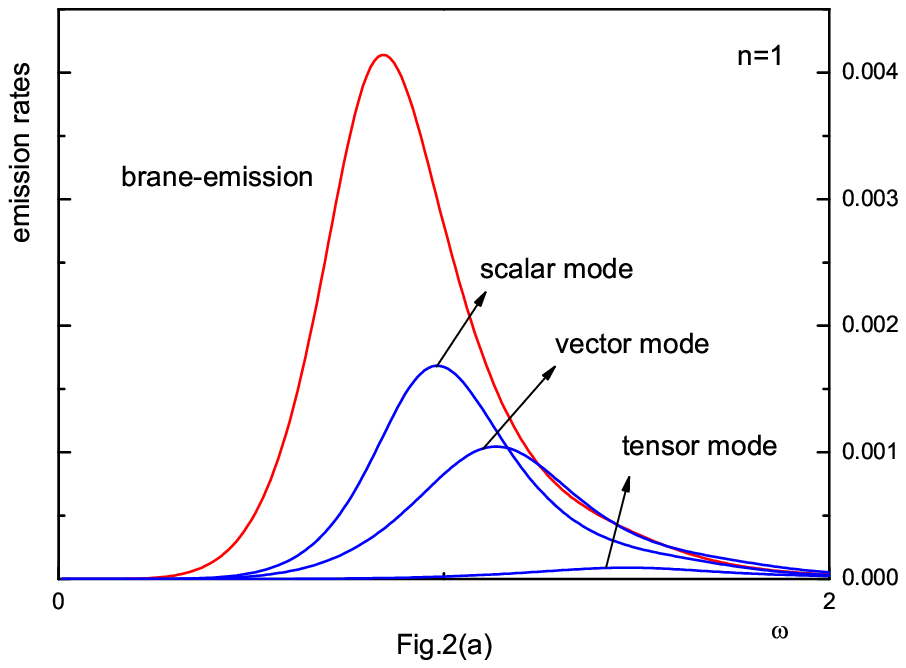}
\includegraphics[height=6cm]{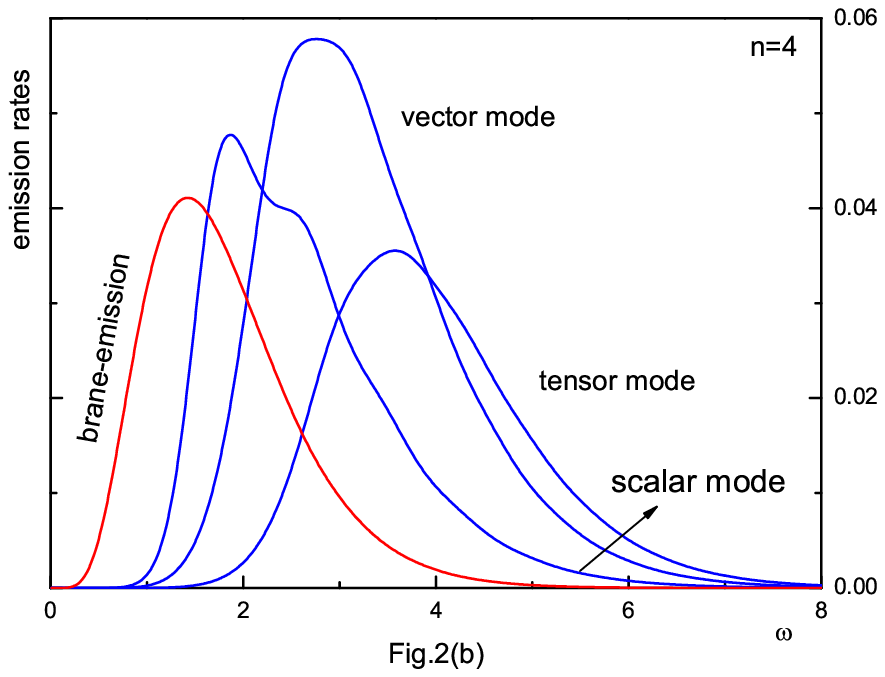}
\caption[fig2]{The $\omega$-dependence of the spin-$2$ field emission spectra for
$n=1$ (Fig. 2(a)) and $n=4$ (Fig. 2(b)). The decay rates on the brane 
are plotted by red color.
The blue line are bulk emission spectra for each mode. This fugure indicates that the
bulk emission rates highly increase with increasing $n$ compared to the brane emission
rates.}
\end{center}
\end{figure}

In Fig. 2 the $\omega$-dependence of the emission spectra is plotted when $n=1$ (Fig. 2(a)) 
and $n=4$ (Fig. 2(b)). The decay rates on the brane 
are plotted by red color. For the comparision
the bulk emission spectra for each mode are plotted together by blue color. Fig. 2 shows 
that the bulk emissivities are in general dominant in the high-energy domain while the brane
decay is dominant in the opposite domain. This is mainly due to the power difference of 
$\omega$ in the emission spectra formula defined in Eq.(\ref{blemi}) and (\ref{bremi}). 
Fig. 2 also indicates that
the bulk decay rates are comparatively larger than the brane decay when $n=4$. 

For the precise comparision we consider the total emission rate defined 
\begin{equation}
\label{totemi}
\Gamma_{tot} \equiv \int_0^{\infty} d \omega \frac{d^2 \Gamma}{d \omega d t}.
\end{equation}

\begin{center}
{\large{Table II}}: Brane versus Bulk in $\Gamma_{tot} / \Gamma_{tot}^S$
\end{center}

\begin{center}
\begin{tabular}{c|c|c|c|c|c|c|c|c}  \hline
\multicolumn{3}{c|} {} & \hspace{.1cm}
 $n=0$ \hspace{.1cm} & \hspace{.1cm} $n=1$ \hspace{.1cm} & \hspace{.1cm} $n=2$
\hspace{.1cm} & \hspace{.1cm} $n=3$ \hspace{.1cm} & \hspace{.1cm} $n=4$
\hspace{.1cm} & \hspace{.1cm} $n=6$  \\ \hline \hline
 {}   & \multicolumn{2}{c|}  {spin-$0$}
& $1$ & $8.82$ & $37.5$ & $99.8$ & $229$ & $784$ \\  \cline{2-9}
Brane & \multicolumn{2}{c|} {spin-$1$}
& $0.45$ & $12.3$ & $65.5$ & $200$ & $463$ & $1600$ \\ \cline{2-9}
{} & \multicolumn{2}{c|} {spin-$2$}
& $0.052$ & $6.82$ & $81.3$ & $167$  & $244$ & $582$  \\ \hline \hline
{} & \multicolumn{2}{c|} {spin-$0$}
& $1$  &  $3.53$
&  $9.01$  &  $22.0$  &  $55.0$  & $408$ \\ \cline{2-9}
{} & {}  &   \hspace{.1cm}  S   \hspace{.1cm}
& $0.23$ & $3.66$ & $16.0$ & $47.2$ & $122$ & $824$ \\ \cline{3-9}
{} & spin-$1$ &  V
& $0.23$ & $2.42$ & $12.8$ & $52.5$ & $193$ & $2383$  \\ \cline{2-9}
{Bulk} & {} & S & $0.026$ & $2.86$ & $27.2$ & $109$ & $312$ & $1976$
\\ \cline{3-9}
{} & spin-$2$ & V & $0.026$ & $2.13$ & $21.4$ & $111$ & $438$ & $5117$
\\ \cline{3-9}
{} & {} & T & {}  & $0.22$ & $4.98$ & $45.9$ & $286$ & $6837$ \\ \hline
\end{tabular}
\\
\end{center}
\vspace{0.5cm}
The relative total emissivities for spin-$0$, spin-$1$, and 
spin-$2$ fields are summarized in Table II. 
Each total emission rate is 
divided by the four-dimensional scalar rate $\Gamma_{tot}^S = 2.98 \times 10^{-4}$. 
The abbreviations S, V and T denote the scalar, vector and tensor modes for the 
bulk fields respectively.

Table II shows several interesting features. Firstly, the total bulk emissivities for the 
spi-$2$ fields become dominant when $n \geq 3$ compared to the emission rate for the 
spin-$2$ fields propagating on the brane.
Secondly, the emission of the spin-$2$ fields into the bulk becomes 
dominant in the presence of the extra dimensions compared to other bulk 
SM fields. In the brane case, however,
the emission rates for the spin-$2$ fields are not dominant. This seems to 
be due to the fact that spin-$2$ graviton, in general,  is not confined 
on the brane unlike the SM
particles.

\begin{center}
{\large{Table III}}: Bulk versus Brane per degree of freedom
\end{center}

\begin{center}
\begin{tabular}{c|cccccc}           \hline
$n$ & \hspace{.1cm} $0$ \hspace{.1cm} & \hspace{.1cm} $1$ \hspace{.1cm} &
\hspace{.1cm} $2$ \hspace{.1cm} & \hspace{.1cm} $3$ \hspace{.1cm} & \hspace{.1cm} $4$
\hspace{.1cm} & \hspace{.1cm} $6$  \\  \hline \hline
spin-$0$ \hspace{.1cm} & \hspace{.1cm} $1.0$ \hspace{.1cm} & \hspace{.1cm} $0.40$
\hspace{.1cm} & \hspace{.1cm} $0.24$ \hspace{.1cm} & \hspace{.1cm} $0.22$ \hspace{.1cm} &
\hspace{.1cm} $0.24$ \hspace{.1cm}
& \hspace{.1cm} $0.52$ \\ 
spin-$1$ \hspace{.1cm} & \hspace{.1cm} $1.0$ \hspace{.1cm} & \hspace{.1cm} $0.33$
\hspace{.1cm} & \hspace{.1cm} $0.22$ \hspace{.1cm} & \hspace{.1cm} $0.20$ \hspace{.1cm} &
\hspace{.1cm} $0.23$ \hspace{.1cm}
& \hspace{.1cm} $0.50$ \\
spin-$2$ \hspace{.1cm} & \hspace{.1cm} $1.0$ \hspace{.1cm} & \hspace{.1cm} $0.31$
\hspace{.1cm} & \hspace{.1cm} $0.15$ \hspace{.1cm} & \hspace{.1cm} $0.23$ \hspace{.1cm} &
\hspace{.1cm} $0.42$ \hspace{.1cm}
& \hspace{.1cm} $1.37$ \\
\hline
\end{tabular}
\end{center}

Table III shows the spin-dependence of the bulk-to-brane emissivities per
degree of freedom(d.o.f.). Since the massless spin-$1$ and spin-$2$ fields have 
$n+2$ and $(n+4) (n+1) / 2$ polarization states respectively, the relative 
emissivities per d.o.f. can be read directly from Table II. Table III indicates that the 
relative emissivities per d.o.f. are always less than unity except spin-$2$ case with
$n=6$, which supports the EHM argument. 

In this paper the emission rates for the spin-$2$ fields on the brane 
and in the bulk are explicitly 
computed. It is found that although the total bulk emissivities becomes dominant when
$n \geq 3$, the bulk-to-brane relative emissivities per degree of freedom remains
$O(1)$, which strongly supports the EHM argument. 
However, as indicated by Table II, the total missing energy arising due to the bulk
emissivities is not negligible. Thus we should carefully consider the missing energy 
portion in the future experiment relating to the blane-world black holes.

It is of interest to 
derive a real effective potentials from the master equation (\ref{master}) by employing the 
transformation theory\cite{chandra92} and interpret the results of the present letter in 
terms of the potentials. It is of greatly interest also to explore the Hawking radiation
for the graviton in the rotating black hole background which is a still open problem. 
We would like to study these issues
in the future.

%\vspace{1cm}
%
{\bf Acknowledgement}:  
This work was supported by the Kyungnam University
Research Fund, 2006.

\end{document}